\documentclass[10pt, conference, compsocconf]{IEEEtran}
\usepackage{nopageno}
\usepackage[T1]{fontenc}
\usepackage{fourier}
\usepackage{float}
\usepackage[english]{babel}							
\usepackage[protrusion=true,expansion=true]{microtype}	
\usepackage{amsmath,amsfonts,amsthm} 
\usepackage[pdftex]{graphicx}	
\usepackage{url}
\usepackage{syntax}
\usepackage{array}
\usepackage{balance}
\usepackage{txfonts}
\usepackage{listings}

\usepackage{fancyhdr}
\numberwithin{equation}{section}		
\numberwithin{figure}{section}			
\numberwithin{table}{section}				

\title{Microservice-based IoT for Smart Buildings}

\author{
\IEEEauthorblockN{Dilshat Salikhov, Kevin Khanda, Kamill Gusmanov \\ Manuel Mazzara, Nikolaos Mavridis}
    \IEEEauthorblockA{Innopolis University, Russia
    \\\{d.salikhov, k.khanda, k.gusmanov, m.mazzara, n.mavridis\}@innopolis.ru}
   
}
\begin{document}
\balance
\maketitle

\begin{abstract}
A large percentage of buildings in domestic or special-purpose is expected to become increasingly "smarter" in the future, due to the immense benefits in terms of energy saving, safety, flexibility, and comfort, that relevant new technologies offer. As concerns hardware, software, or platform level, however, no clearly dominant standards currently exist. Such standards, would ideally, fulfill a number of important desiderata, which are to be touched upon in this paper. Here, we will present a prototype platform for supporting multiple concurrent applications for smart buildings, which is utilizing an advanced sensor network as well as a distributed microservices architecture, centrally featuring the Jolie programming language. The architecture and benefits of our system are discussed, as well as a prototype containing a number of nodes and a user interface, deployed in a real-world academic building environment. Our results illustrate the promising nature of our approach, as well as open avenues for future work towards its wider and larger scale applicability.
\end{abstract}

\section{Introduction}
The Internet of Things (IoT), as per the ITU Recommendation ITU-T Y.2060 \cite{IoT}, has been defined as: \textit{"a global infrastructure for the information society, enabling advanced services by interconnecting (physical and virtual) things based on existing and evolving interoperable information and communication technologies."} Among physical objects, buildings are playing a major role in this technological transition. Building function either as human habitats, for example domestic buildings, or for some specialized goals, for example storehouses, shops, industrial buildings, or schools. In some cases they can have both functions.

There are multiple aspects of the operations of modern-day buildings that allow for further automation and optimization. It has been shown that the benefits of an improved energy management through automation are indeed significant \cite{Nunes2017}; also, security of buildings, human-friendliness and adaptation to preferences has a vast spectrum of improvement. \textit{Smart buildings}, as well as \textit{smart cities} that are supposed to contain them, are the target of much research today, and promise to dramatically improve our lives increasing sustainability and improving the environment.

Traditionally, most building automation systems were made for specific applications and offered little degree of openness and flexibility. However, with the fast maturation of a number of supporting technologies, the opportunity to change this \textit{status quo} is rapidly growing. First, cheap sensing and perception technologies have become available for a wide range of measurables: covering not only physical properties of the building and its spaces, such as temperature, light, and humidity, but also providing information about the presence, number, identities, activities, and even emotional states of the people inside a building or in its surrounding spaces
\cite{Nalin2016}. Second, affordable and miniature microprocessor-based platforms have become widespread and are easily inter-connectable to sensors, which often have enough processing power to support perception and machine vision; with multiple network transports, even the necessary small battery power is readily available. Third, networking technology for such platforms has advanced significantly, and nowadays it is easy to implement building-wide networks, often with dynamic ad-hoc topologies, which also support on-the-fly introduction and replacement of new nodes, while providing secure communications. Fourth, special languages and middleware has been developed, in order to support straightforward development of distributed systems based around a number of paradigms, including microservices \cite{Dragoni16,Mazzara2016}

Given all these developments, one cannot only envision -- but can also start implementing and experimenting with -- the usage of the IoT for providing generic smart building infrastructure, which goes beyond the constraints of existing specialized systems, and provides fluid support for diverse concurrent applications, sharing the infrastructure, and operating through microservices provided by the nodes. Furthermore, infrastructure such as the \textit{"Jolie Good Buildings"} that we present here, promises strong scalability, reliability, and upgradability as more powerful hardware becomes available, and finally direct utilization of the immense power of external services available on the Internet, as part of the distributed applications running on the nodes. 

In this paper, we will start by providing background on relevant existing work. Then, we will present the overall architecture of our system and will explain the requirements and design choices we have made. We will then present results, discuss current and future steps, and finish with a forward-looking conclusion. We hope that this work will help create a future, in which not only resources are saved and the environment protected, but also human life becomes less stressful with enhanced efficiency and creativity. 

\section{Background}
The IoT is an important component of current and future human life. At the moment, the number of connected devices is greater than the estimate of 22 billion \cite{devices}, which is several times greater than the number of earth's inhabitants. However, the problem arises that many devices are incompatible with each other, while others are used for other purposes. What kind of information can you get from one fitness tracker? Using the accelerometer and gyroscope, you can get information about what people are now running or just walking, eating or swimming in the pool. However, the most important issue is the ability to work with this information. On the basis of data from the heart rate sensor, you can get a small feature on the state of human health and if required, suggest various treatments. Also, this information is very useful in medicine, the physician can better evaluate diseases based on previous data, for example when the development began, how often problems happened and so on.

\subsection{Smart Building and Cloud}
The IoT can be used in almost all spheres of human life. In this paper, we focus on Smart Buildings and Human-Building interactions. There have been many different studies on the topic, which uses different kinds of sensors, frameworks and, sometimes, robots. The creation of a single and simple platform that can communicate with all types of sensors and robots is a very important step towards the development of Smart Buildings. It should be noted that it is impossible to use large computers for work in each building. In this regard, there is need for Cloud Computing (CC) where computation is viewed as a "utility". In a similar sense, with modern power and water networks, cloud users do not need to own the means of production or distribution (i.e. power generators, water sources and distribution networks): they just connect to the cloud, and time-share reusable distant distributed computation, storage, and code resources in a transparent fashion (not knowing the whereabouts or the specifics of them) and with high robustness. 

Yet, CC also has a number of limitations. For example, it is not effectively connected to the physical world in the way that a situated robotic agent would be \cite{Mavridis2012}. In this regard, there is the problem of creating a service view of a Human-Robot Cloud, in our case a Human-Sensor Cloud. A further stage of development of the system of Smart Buildings is to create a system of interaction between sensors and robots in buildings and humans. Frameworks for human-machine systems \cite{Mavridis2013} are created and used as the basis for interaction with robots, and provide easy and fast connection with new devices instead of old ones. In fact, the connection and further work with new devices strongly affects the level of performance. Creating a platform for connection and operation of devices of various types is necessary to build the right network.

\subsection{Microservices}
Microservices are very useful for the IoT. The Microservice software architecture is a style for service-oriented development whose popularity is increasing now because of growing interest to parallel computations. Designed and built for microservices architecture, Jolie programming language was developed to work directly with a service-oriented paradigm, which distinguishes it from other popular languages like C\#, Java, or Python. This means that the language contains features that are unique to this approach, i.e. example representation of building blocks. In object-oriented languages, there are usually classes or functions; in Jolie, building blocks are a service in their own right.

The Jolie programming language \cite{jolie,jolie:website} was created
in order to maximize the use of the microservices architectural style. In Jolie, microservices are first-class citizens: every microservice can be reused, orchestrated, and aggregated with others~\cite{montesi}. This approach brings simplicity in components management, reducing development and maintenance costs, and supporting distributed deployments~\cite{fowler-tradeoffs}.

\begin{figure*}[h]
\centering{\includegraphics[scale=0.8]{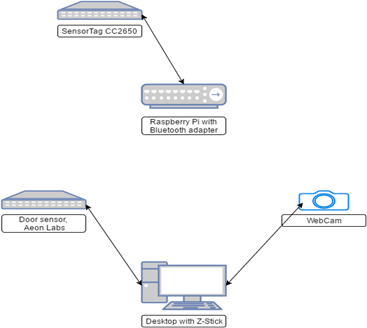}}
\caption{A scheme of devices and their connections}
\label{fig:arch}
\end{figure*}

The development of Jolie followed a major formalization effort for workflow and service composition languages, and the EU Project SENSORIA~\cite{sensoria} has successfully produced a plethora of models for reasoning about composition of services (e.g.,~\cite{LucchiM07,Mazzara11,mazzaraPhD}). On the mathematical side, the formal semantics of Jolie~\cite{sock,GLMZ09,MC11} have been inspired
by different process calculi, such as CCS \cite{CCS:Milner80} and the $\pi$-calculus~\cite{MPW92}. From a practical point of view, however, Jolie is a descendant of standards for Service-Oriented Computing such as, for example, WS-BPEL~\cite{bpel}. With both theoretical and practical influences, Jolie is a suitable candidate for the application of recent research techniques, e.g., runtime adaptation~\cite{PGLMG14}, process-aware web applications~\cite{M14},or correctness-by-construction in concurrent software~\cite{CM13}.

\section{Architecture}

At this stage, the main goal is to create a small and simple system, where a major responsibility was taken by Jolie \ref{fig:arch}. In order to achieve the goal the entire process was divided in steps.

\subsection{Step 1: Connecting and configuring sensors}
The first step was connecting and configuring sensors through BLE. The main data collected for temperature, humidity and luminosity were made using CC2650 SensorTag because they have the necessary sensors, small size, run on batteries, and easy code to write. The work with sensors are very similar to the creation of the sketches for Arduino: C code is written with the necessary libraries for the SensorTag, sensors to be used, and ways of exchanging information with these devices. In our case, data is sent via BLE, as it is very easy to use and configure.
The door sensor from Aeon Labs, which works on the protocol Z-Wave, was used to create a monitoring system entrance/exit to the premises. To work with this device, HomeOS \cite{Mavridis2015} was used, which is written in the programming language C\# and has the necessary functionality to work with the devices of this company in the protocol Z-Wave.

\begin{figure*}[h]
\centering{\includegraphics[scale=0.25]{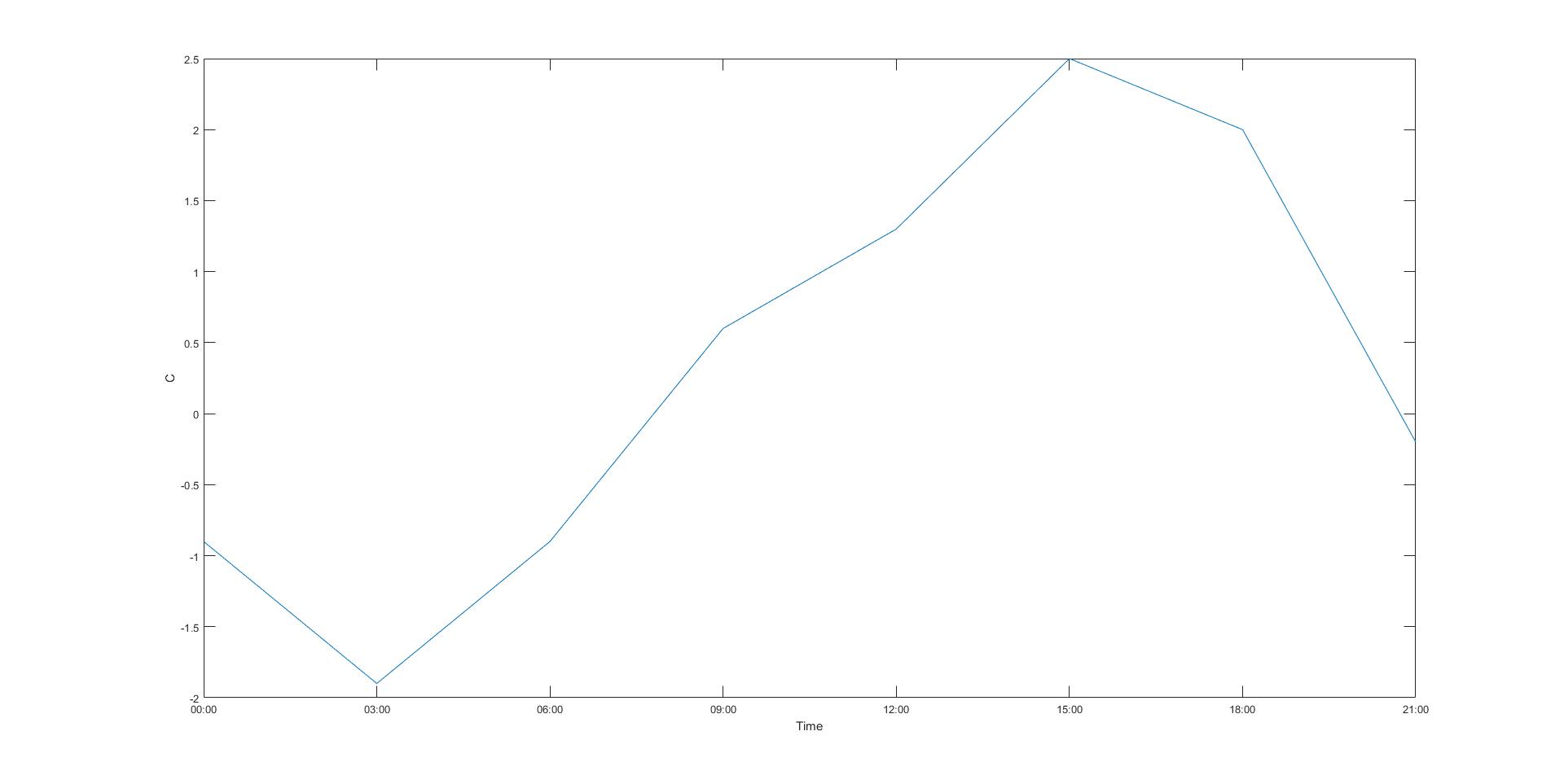}}
\caption{A graph of the outdoor temperature. Axis of abscissae: Time of the day. Axis of ordinates: Celsius degrees}
\label{fig:weather}
\end{figure*}

\subsection{Step 2: Coding in Jolie to work with sensors}
The next step was to writing simple code to work with the sensors in Jolie. A major advantage of this approach is code reusability. Our system will support different types of sensors snf the main logic of their connectivity and data extraction is similar for most of them. Also, product support becomes easier (reusability). The same service can be used for these sensors. Implied by the first advantage, the second advantage is reducing bugs. A third advantage that may be interesting for future collaborators is simplicity. Jolie divides all system logic into small parts: we have several services that are responsible for each sensor or each action, the naming of each block is intuitively understandable, so these language features increase code readability. In addition, one more significant advantage is working with Java code in Jolie. Thus, the part of the work with getting data from BLE devices was written in the Java programming language and divided into simple functions, e.g. connecting to devices and data retrieval, which is called from Jolie. Thus, the code is easy to read, clear, and ready for further work as a client for other sensors.

\subsection{Step 3: Data Collection}
The last step is data collection. From SensorTags, we read data about room temperature, humidity, light and pressure. Also we parsed outdoor temperature from Internet~\ref{fig:weather}. All of them were recorded in .csv format for later processing and graphing in MATLAB. Data from Aeon Labs sensor is the number of opening/closing doors and monitoring incoming/outgoing people using the webcam, which recorded in .csv and .jpg format respectively. It is also worth noting that all codes worked in Raspberry Pi, which has all necessary libraries and connections.
Despite its apparent simplicity, there were several issues at work with devices. The first serious problem was the lack of libraries for Java to work with BLE devices. But we found a great and simple library for Intel Edison devices and used it. The next issue was working with Z-Wave devices in HomeOS, which periodically generated exceptions and did not connect to the device. However, all these problems were solved, which allowed us to come to the result, which will be described in the next section.

\section{Results}
We have developed a platform to work with the SensorTags based on Jolie and Java programming languages. Despite its small size, it obtained most of the data, with which we work. The ability to embed Java code allows the possibility to add necessary functions and give more functionality and usability to Jolie language. At this stage, our goal was to build the basic functionality of working with sensors, understanding how they work and how to interact with them and get ideas for the further development of the project. In order to analyze performances, we have placed sensors in three rooms: the students' rooms in the dormitory and the laboratory of Innopolis University (Russian Federation), in which we spend most time of the day. It is worth noting that both rooms can be ventilated and fitted with lights and heaters. 


\begin{figure*}[h]
\centering{\includegraphics[scale=0.25]{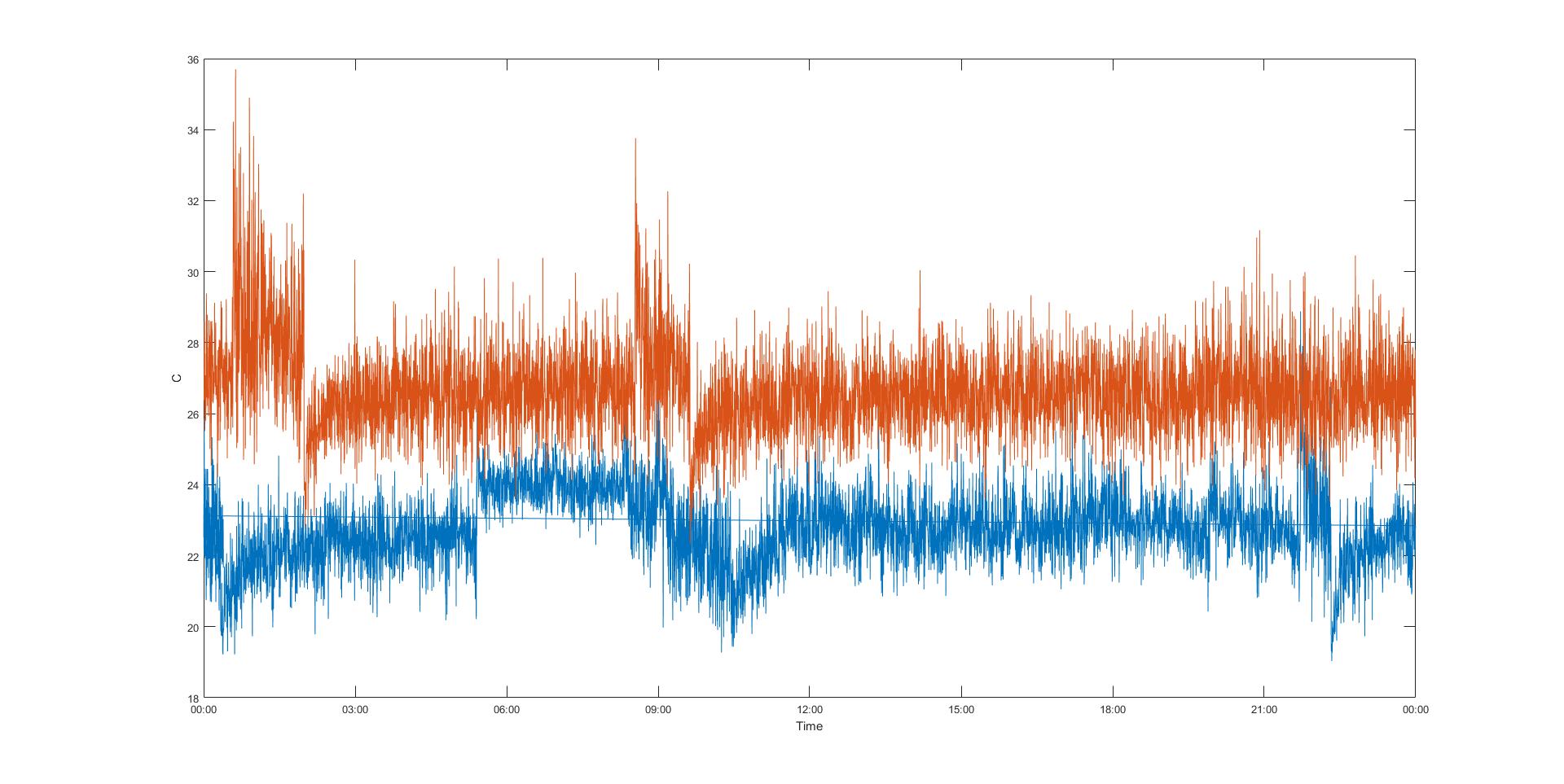}}
\caption{Temperature in the rooms. Blue figure in first room and red in second. Axis of abscissae: Time of the day. Axis of ordinates: Celsius degrees}
\label{fig:temp1}
\end{figure*}

\begin{figure*}[h]
\centering{\includegraphics[scale=0.3]{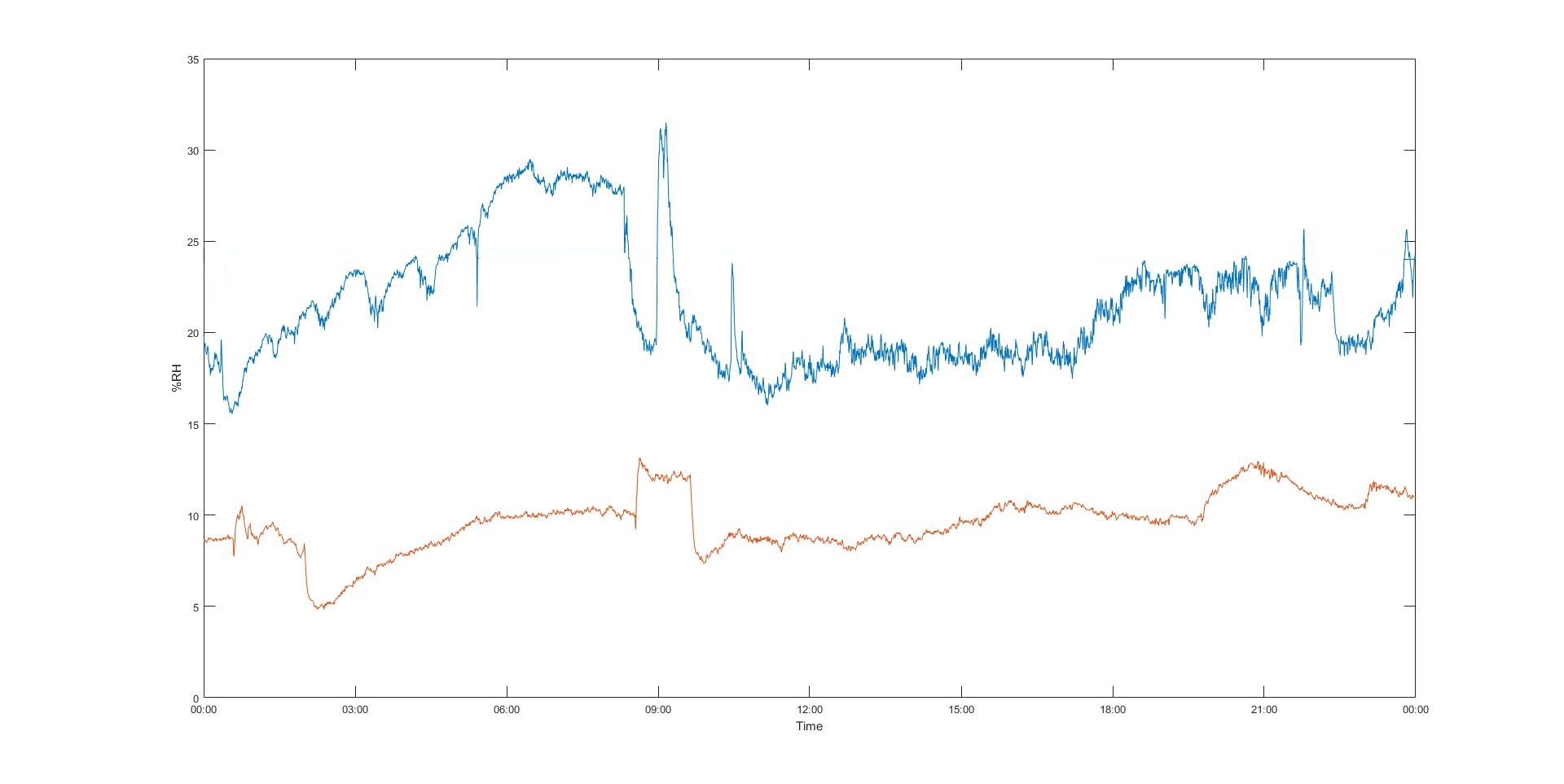}}
\caption{Humidity in the room. Axis of abscissae: Time of the day. Axis of ordinates: Percentage of Relative humidity (RH)}
\label{fig:hum}
\end{figure*}

\begin{figure*}[h]
\centering{\includegraphics[scale=0.3]{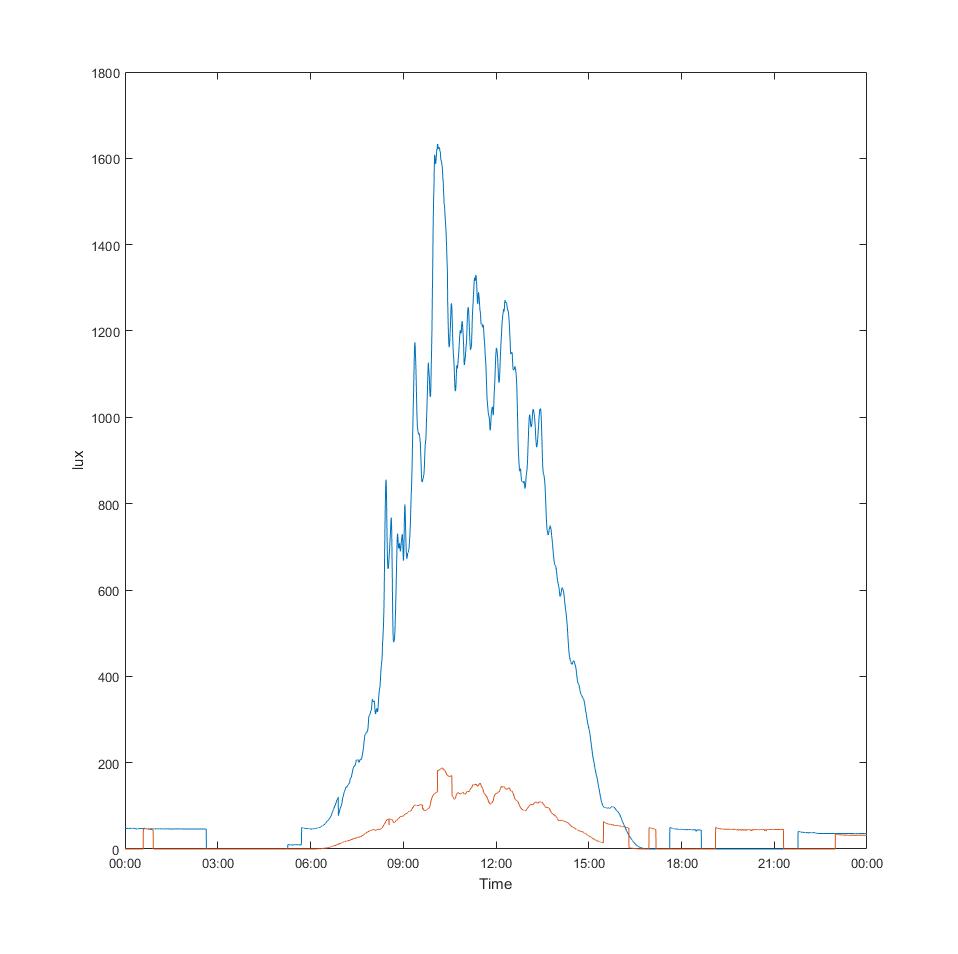}}
\caption{A graph of the data from light sensor in the rooms. Axis of abscissae: Time of the day. Axis of ordinates: Lux}
\label{fig:light1}
\end{figure*}

\begin{figure*}[h]
\centering{\includegraphics[scale=0.3]{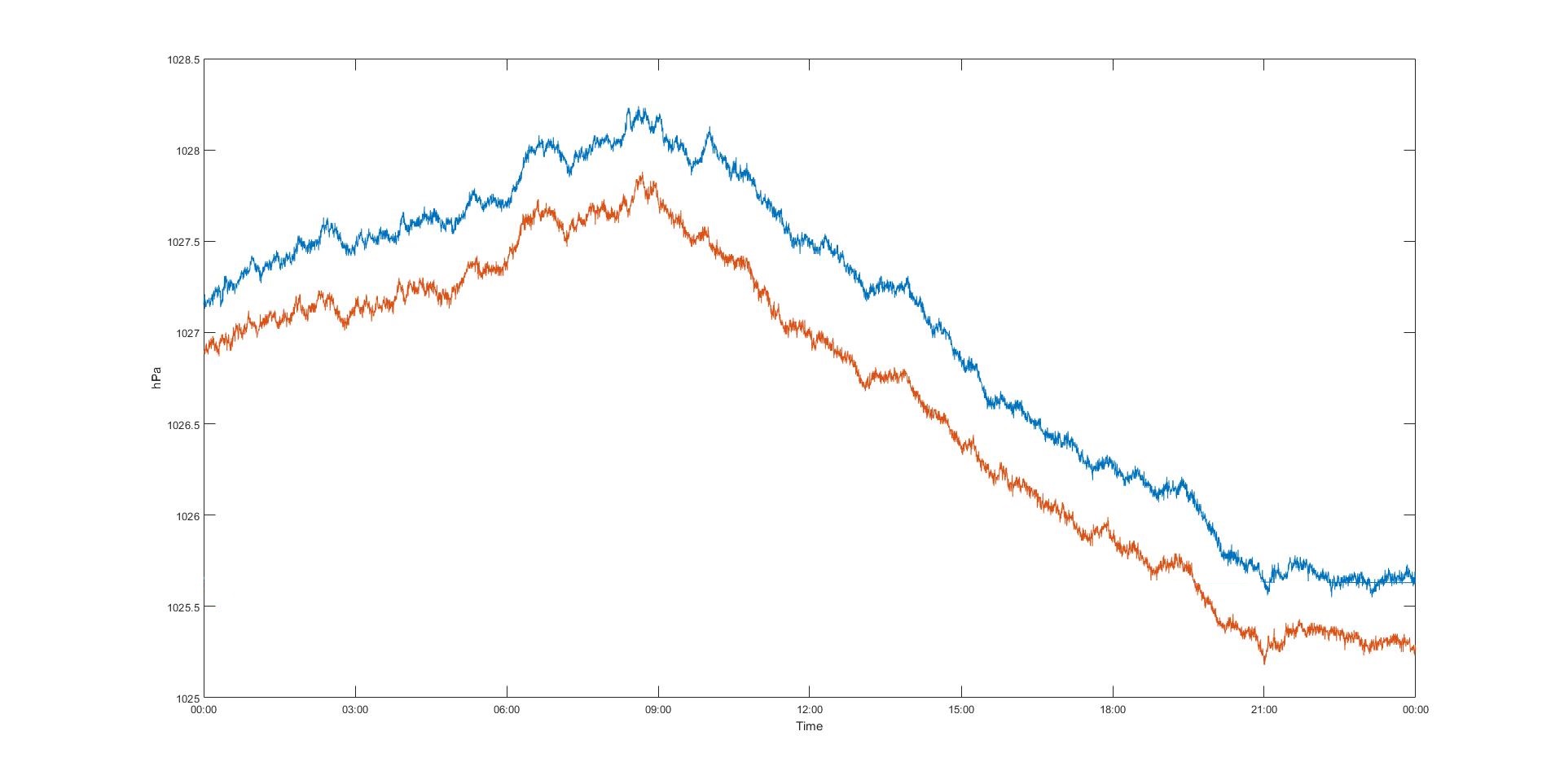}}
\caption{A graph of the pressure in the rooms. Axis of abscissae: Time of the day. Axis of ordinates: millibars}
\label{fig:pressure1}
\end{figure*}

In the students' room we set the Raspberry Pi with the connected Bluetooth adapter and the SensorTag to obtain data on the temperature. For tracking of location of the student in the room, we used data from a fitness-tracker MiBand2, using its mac address. Since the room is small we could accurately determine when the student enters and when he/she leaves. The collected dataset has been used to construct graps of temperature \ref{fig:temp1}, humidity \ref{fig:hum}, light \ref{fig:light1} and pressure \ref{fig:pressure1} in the room, which will later be used optimize environment conditions.

Regarding humidity trend, the ideal value should be 40-50\% ~\cite{relhumidity}. In our experiment humidity is always lower than 30\%, which causes discomfort and can lead to illness, and more in general does not support students to perform at their best in daily intellectual activities. 

For the light sensor it is worth noticing that in the second room values are never greater than 200 lux, while in first room there is an average of 600 lux. The reason is that the second room is located between two building and never receive sufficient sunlight, while the first has an open view from a luminous window.

In the laboratory, we placed more devices and used multiple platforms to work with them. This time SensorTag also sent data about room temperature, humidity, light and pressure. Also, we used a sensor of opening/closing doors and a web camera to record those entered/exited the room. As noted earlier, for obtaining data on temperature, humidity and light, their processing and saving to file, we used two programming language: Java and Jolie, which were running on the Raspberry Pi. Work with other sensors and a camera was assigned to HomeOS, which has the necessary functionality for working with the devices.Using a sensor on the door and the camera data, we obtained data how many people were in the room in each moment of time and assembled a small dataset of lab workers photos. This completes the main work of the first stage. Plans for further development of the project and data will be described in the next section.

\section{Future Steps}
Despite of having solved many of the issues related to obtaining data from the sensors and its further processing, in order to get the targeted result there are further aspects that need to be taken into consideration. First, we must configure all SensorTags to work on the ZigBee protocol. In the future we will set up these sensors in a variety of classrooms at our university, and there is a need to build a mesh network for communication between devices and further processing in the language Jolie. Second, we need to expand the functionality of the language Jolie to work with all the devices. In this case, we will have to write the modules for BLE, ZigBee and Z-Wave devices to continue to connect easily and share data between smart devices. In addition, we want to rewrite the module for working with the Aeon Labs sensors in Jolie language. 

Finally, we need to place all sensors and the Raspberry Pi in the classroom of the University and in the room of the student and start collecting data. In the room of the student, they will continue collecting data on temperature including the data on humidity and lighting as well as installing a window-opening sensor that will allow us to get good dataset to determine the preferred mode of temperature, light and humidity in the room and drawing a small daily schedule of a student. In the classrooms, we will also install a small camera to track the number of people in the room and launch a small bot to Telegram messenger to gather feedback from students in terms of comfort temperature and humidity in the auditorium. In the lab, we will add a motion sensor and several relays to control lights and power sockets in the room that will be activated either automatically or manually through a central control system. 

Based on the obtained data, which will take about a month to collect, we will built a prediction system for students and employees, which will be checked by a simple small bot in Telegram. Thus, we will improve our result in order to  create a small artificial intelligence responsible for the control of lighting, temperature, electricity and comfort level in the room. 

This will lead to the creation of a unified platform, unified control center, written in Jolie that will allow:

\begin{enumerate}
\item To develop a system of microservices in Jolie
\item Easy to incorporate new devices operating according to one of three protocols
\item To track the preferences of students and staff in the environment
\item To control the flow of electricity to the premises and, if possible, to suggest ways to reduce costs
\item Develop the idea of Human-Building Interaction and technologies on campus and the entire city;
\end{enumerate}

On the software engineering side, instead, all the development process has to be streamlined and organized in a more formal way, from requirements elicitation to deploy and testing. Formal techniques will be here of major help \cite{Mazzara2010,Mazzara2011}.

\section{Conclusions}
In this paper we presented preliminary steps toward the transformation of the Innopolis University building and students dorms into an effective smart building. Tracking some key environmental values emphasized the need for an optimization to lead to both to energy save and improved living conditions.
In the future it is not impossible to imagine buildings capable to adapt and self-configure depending on environmental conditions and human needs, in the same way as modern software shows the same flexibility \cite{MDZ2011}.

Jolie demonstrated to be flexible and simple enough for working with microservices and the Internet of Things: code is easy to write, to deploy and devices are easy t connect. The overall scenario looks promising to be repicated in several projects related to smart homes and cities. 

A large obstacle to the development is existence of a narrow set of tools for the job, and limited documentation, that will require specific attention on a case to case basis. However, the Internet community and students of many European Universities supported us in our enterprise. In the future, this work will allow us to create a system of HBI that is easier to use and more affordable. 

\section*{Acknowledgements}
The authors received logistic and financial support by Innopolis University. We would like to thank Daniel Johnston for linguistic assistance and all the colleagues and students who made this work possible.

\bibliographystyle{plain}
\bibliography{IoT}

\begin{thebibliography}{10}

\bibitem{sensoria}
{EU Project SENSORIA. Accessed April 2016}.
\newblock \url{http://www.sensoria-ist.eu/}.

\bibitem{IoT}
Internet of things global standards initiative.
\newblock \url{http://www.itu.int/en/ITU-T/gsi/iot/Pages/default.aspx}.
\newblock Online; accessed 25th of October 2016.

\bibitem{devices}
Internet of things (iot): number of connected devices worldwide from 2012 to
  2020.
\newblock
  \url{https://www.statista.com/statistics/471264/iot-number-of-connected-devices-worldwide/}.
\newblock Online; accessed 25th of October 2016.

\bibitem{jolie:website}
{Jolie Programming Language. Accessed April 2016.}
\newblock \url{http://www.jolie-lang.org/}.

\bibitem{relhumidity}
Relative humidity and your home.
\newblock
  \url{http://www.thermastor.com/information/relative-humidity-and-your-home.aspx}.
\newblock Online; accessed 25th of October 2016.

\bibitem{bpel}
{WS-BPEL} {OASIS} {W}eb {S}ervices {B}usiness {P}rocess {E}xecution {L}anguage.
  accessed {April} 2016.
\newblock
  \url{http://docs.oasis-open.org/wsbpel/2.0/wsbpel-specification-draft.html}.

\bibitem{CM13}
Marco Carbone and Fabrizio Montesi.
\newblock Deadlock-freedom-by-design: multiparty asynchronous global
  programming.
\newblock In {\em POPL}, pages 263--274, 2013.

\bibitem{Dragoni16}
Nicola Dragoni, Saverio Giallorenzo, Alberto Lluch{-}Lafuente, Manuel Mazzara,
  Fabrizio Montesi, Ruslan Mustafin, and Larisa Safina.
\newblock Microservices: yesterday, today, and tomorrow.
\newblock {\em to appear in Present And Ulterior Software Engineers, Springer},
  2016.

\bibitem{fowler-tradeoffs}
M.~Fowler.
\newblock {Microservice Trade-Offs}.
\newblock \url{http://martinfowler.com/articles/microservice-trade-offs.html},
  (2015).

\bibitem{GLMZ09}
C.~Guidi, I.~Lanese, F.~Montesi, and G.~Zavattaro.
\newblock Dynamic error handling in service oriented applications.
\newblock {\em Fundam. Inform.}, 95(1):73--102, 2009.

\bibitem{sock}
C.~Guidi, R.~Lucchi, G.~Zavattaro, N.~Busi, and R.~Gorrieri.
\newblock Sock: a calculus for service oriented computing.
\newblock In {\em ICSOC, volume 4294 of LNCS}, pages 327--338. Springer, 2006.

\bibitem{LucchiM07}
R.~Lucchi and M.~Mazzara.
\newblock A pi-calculus based semantics for {WS-BPEL}.
\newblock {\em J. Log. Algebr. Program.}, 70(1):96--118, 2007.

\bibitem{Mazzara2016}
Larisa Safina Ivan~Lanese Manuel~Mazzara, Ruslan~Mustafin.
\newblock Towards microservices and beyond: An incoming paradigm shift in
  distributed computing.
\newblock {\em arXiv preprint arXiv:1610.01778}, 2016.

\bibitem{Nalin2016}
Ilaria~Baroni Marco~Nalin and Manuel Mazzara.
\newblock A holistic infrastructure to support elderlies' independent living.
\newblock {\em Encyclopedia of E-Health and Telemedicine, IGI Global}, 2016.

\bibitem{Mavridis2012}
Nikolaos Mavridis, Thirimachos Bourlai, and Dimitri Ognibenes.
\newblock The human-robot cloud: Situated collective intelligence on demand.
\newblock {\em Cyber Technology in Automation, Control, and Intelligent Systems
  (CYBER), 2012 IEEE International Conference}.

\bibitem{Mavridis2013}
Nikolaos Mavridis, Stasinos Konstantopoulos, Ioannis~A. Vetsikas, I.~Heldal,
  Pythagoras Karampiperis, G.~Mathiason, Serge Thill, K.~Stathis, and Vangelis
  Karkaletsis.
\newblock {CLIC:} {A} framework for distributed, on-demand, human-machine
  cognitive systems.
\newblock {\em CoRR}, abs/1312.2242, 2013.

\bibitem{Mavridis2015}
Nikolaos Mavridis, Georgios Pierris, Chiraz BenAbdelkader, Aleksandar Krstikj,
  and Christos Karaiskos.
\newblock Smart buildings and the human-machine cloud.
\newblock In {\em GCC Conference and Exhibition (GCCCE), 2015 IEEE 8th. IEEE,
  2015}.

\bibitem{MDZ2011}
Dragoni Nicola Zhou~Mu. Mazzara, Manuel.
\newblock {Dependable workflow reconfiguration in WS-BPEL}.
\newblock In {\em Proceedings of the 5th Nordic Workshop on Dependability and
  Security}, 2011.

\bibitem{Mazzara11}
M.~Mazzara, F.~Abouzaid, N.~Dragoni, and A.~Bhattacharyya.
\newblock Toward design, modelling and analysis of dynamic workflow
  reconfigurations - {A} process algebra perspective.
\newblock In {\em Web Services and Formal Methods - 8th International Workshop,
  {WS-FM}}, pages 64--78, 2011.

\bibitem{mazzaraPhD}
Manuel Mazzara.
\newblock {\em Towards Abstractions for Web Services Composition}.
\newblock PhD thesis, University of Bologna, 2006.

\bibitem{Mazzara2010}
Manuel Mazzara.
\newblock Deriving specifications of dependable systems: toward a method.
\newblock In {\em 12th European Workshop on Dependable Computing (EWDC)}, 2009.

\bibitem{Mazzara2011}
Manuel Mazzara.
\newblock On methods for the formal specification of fault tolerant systems.
\newblock In {\em 4th International Conference on Dependability(DEPEND)}, 2011.

\bibitem{CCS:Milner80}
R.~Robin Milner.
\newblock {\em {A} calculus of communicating systems}.
\newblock Lecture notes in computer science. Springer-Verlag, Berlin, New York,
  1980.

\bibitem{MPW92}
Robin Milner, Joachim Parrow, and David Walker.
\newblock A calculus of mobile processes, {I and II}.
\newblock {\em Information and Computation}, 100(1):1--40,41--77, September
  1992.

\bibitem{MC11}
F.~Montesi and M.~Carbone.
\newblock {Programming Services with Correlation Sets}.
\newblock In {\em Proc. of Service-Oriented Computing - 9th International
  Conference, {ICSOC}}, pages 125--141, 2011.

\bibitem{montesi}
Fabrizio Montesi.
\newblock {J}{O}{L}{I}{E}: a {S}ervice-oriented {P}rogramming {L}anguage.
\newblock Master's thesis, University of Bologna, 2010.

\bibitem{M14}
Fabrizio Montesi.
\newblock {Process-aware web programming with Jolie}.
\newblock In {\em Proceedings of the 28th Annual ACM Symposium on Applied
  Computing}, SAC '13, pages 761--763, New York, NY, USA, 2013. ACM.

\bibitem{jolie}
Fabrizio Montesi, Claudio Guidi, and Gianluigi Zavattaro.
\newblock Service-oriented programming with jolie.
\newblock In {\em Web Services Foundations}, pages 81--107. 2014.

\bibitem{Nunes2017}
Renato Jorge~Caleira Nunes.
\newblock Home automation - a step towards better energy management.
\newblock In {\em INTERNATIONAL CONFERENCE ON RENEWABLE ENERGIES AND POWER
  QUALITY (ICREPQ 2017)}.

\bibitem{PGLMG14}
Mila~Dalla Preda, Saverio Giallorenzo, Ivan Lanese, Jacopo Mauro, and Maurizio
  Gabbrielli.
\newblock {AIOCJ:} {A} choreographic framework for safe adaptive distributed
  applications.
\newblock In {\em Software Language Engineering - 7th International Conference,
  {SLE} 2014, V{\"{a}}ster{\aa}s, Sweden, September 15-16, 2014. Proceedings},
  pages 161--170, 2014.

\end{thebibliography}

\end{document}